\documentclass{emulateapj}

\newcommand{\ltarget}{SDSSJ125733.63+542850.5}
\newcommand{\target}{SDSSJ1257+5428}
\newcommand{\kms}{\mbox{km}\,\mbox{s}^{-1}}
\newcommand{\msun}{\mbox{M}_\odot}

\begin{document}

\title{Detection of a white dwarf companion to the white dwarf 
  \ltarget\footnote{Based on observations made with the William Herschel
    Telescope operated on the island of La Palma by the Isaac Newton Group in
    the Spanish Observatorio del Roque de los Muchachos of the Instituto de 
    Astrof\'{i}sica de Canarias.}
}

\author{T.R. Marsh, B.T. G\"ansicke, D. Steeghs}
\affil{Department of Physics, University of Warwick, Coventry CV4 7AL,
  UK}
\email{t.r.marsh@warwick.ac.uk}
\author{J. Southworth}
\affil{Astrophysics Group, Keele University, Keele, Staffordshire ST5
  5BG, UK}
\author{D. Koester}
\affil{Institut f\"ur Theoretische Physik und Astrophysik, University of Kiel,
24098 Kiel, Germany}
\author{V. Harris}
\affil{133 Jackman Close,  Abingdon, Oxfordshire OX14 3GB, UK}
\author{L. Merry}
\affil{Flat 4, 8 The Lakes, Larkfield, Kent ME20 6GE, UK}


\begin{abstract}
\ltarget\ (hereafter \target) is a compact white dwarf binary from the Sloan
Digital Sky Survey that exhibits high-amplitude radial velocity variations on
a period of $4.56\,$hours. While an initial analysis suggested the presence of
a neutron star or black-hole binary companion, a follow-up study concluded
that the spectrum was better understood as a combination of two white dwarfs.
Here we present optical spectroscopy and ultraviolet fluxes which directly
reveal the presence of the second white dwarf in the system.  \target's
spectrum is a composite, dominated by the narrow-lined spectrum from a cool,
low gravity white dwarf ($T_{\mathrm{eff}} \simeq 6300\,$K, $\log g = 5$ to
$6.6$) with broad wings from a hotter, high-mass white dwarf companion
($11,000$ to $14,000\,$K; $\sim 1\,\msun$). The high-mass white dwarf has
unusual line profiles which lack the narrow central core to H$\alpha$ that is
usually seen in white dwarfs. This is consistent with rapid rotation with $v
\sin i = 500$ to $1750\,\kms$, although other broadening mechanisms such as
magnetic fields, pulsations or a helium-rich atmosphere could also be
contributory factors. The cool component is a puzzle since no evolutionary
model matches its combination of low gravity and temperature.  Within the
constraints set by our data, \target\ could have a total mass greater than the
Chandrasekhar limit and thus be a potential Type~Ia supernova
progenitor. However, \target's unusually low mass ratio $q \approx 0.2$
suggests that it is more likely that it will evolve into an accreting double
white dwarf (AM~CVn star).
\end{abstract}

\section{Introduction}

Close pairs of white dwarfs with combined masses greater than the
Chandrasekhar limit have long been discussed as potential progenitors of
Type~Ia supernovae \citep{IbenTutukov1984,Webbink:DDs}. However, despite
searches, \citep{Robinson:DDs,Bragaglia:DDs,Marsh:friends,Napiwotzki:SPY} no
secure examples of double white dwarfs both massive enough and short-period
enough to merge within a Hubble time have been found. Only about 1 in a 1000
white dwarfs are required to be in such systems to match Type~Ia rates
\citep{Nelemans:DWDs}. Since only of order 1000 white dwarfs have been
searched for binarity to date, the deficit is not yet significant, however it
continues to be worth searching for more such systems. In an effort to do so
we examined the spectra of DA white dwarfs (those showing spectra with
hydrogen absorption only) from the SDSS survey \citep{Eisenstein:wds}, looking
for objects of discrepant radial velocity.  One star, \target\ was an obvious
outlier with a mean radial velocity of $-300\,\kms$. In this paper we present
follow-up spectroscopy to elucidate the nature of this object.

\target\ was the subject of a similar study by
\cite{Badenes:SDSS1257}. They found that it was a binary, measured a radial
velocity semi-amplitude of $K_1 = 323 \pm 6\,\kms$ on a period of
$4.56\,$hours and fitted their spectra with a white dwarf of temperature $\sim
9000\,$K and high mass, $\sim 0.9\,\msun$. The period, radial velocity
amplitude and white dwarf mass give a minimum mass for the companion of
$1.6\,\msun$, suggesting that it is either a neutron star or
black-hole. \cite{Badenes:SDSS1257} estimated a distance of $48\,$pc for
\target, implying that such systems may be rather common.

\begin{figure*}
 \centering
\hspace*{\fill}
\includegraphics[width=0.9\textwidth]{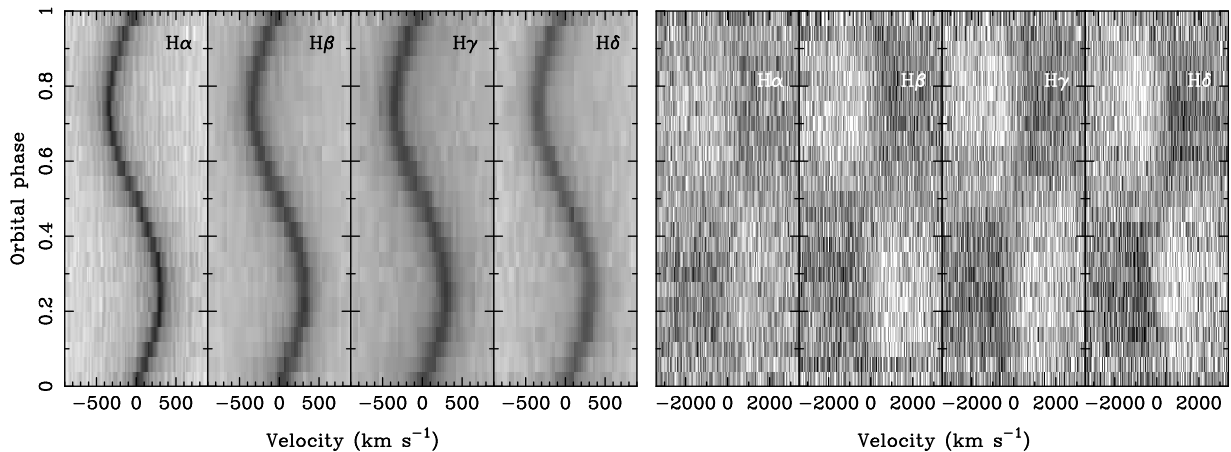}
\hspace*{\fill}
\caption{The panels show phase-folded trailed spectra of the first four
  Balmer lines of \target, H$\alpha$--$\delta$, left to right. The
  left-hand panel shows the raw data; the right-hand panel, which shows the
  data after removal of the primary's motion and mean spectrum, reveals anti-phased
  features from the massive secondary white dwarf. (Note the change of
  horizontal scale between the two panels.)
\label{fig:pbin_trail}}
\end{figure*}

\cite{Badenes:SDSS1257} recognised that their spectral fits were problematic,
and in particular failed to fit the narrow cores of the Balmer lines. Thus
while they favored a neutron star or a black-hole for the companion, they
could not entirely eliminate the possibility that it was another white dwarf.
\cite{kulkarni+vankerkwijk10-1} took three high signal-to-noise spectra which
showed asymmetries suggesting exactly this; they showed that their spectra
could be fit by a combination of a cool, low mass white dwarf 
($6250 \pm 250\,$K, $0.15\pm 0.05\,\msun$) plus a hotter, massive white dwarf 
companion ($13$,$000\pm 800\,$K, $0.92 \pm 0.13\,\msun$).

\cite{kulkarni+vankerkwijk10-1}'s paper appeared shortly after the original
submission of our work. Although both our papers agreed upon the basic
double white dwarf nature of \target, there were significant differences of
detail, with inconsistencies in both masses and temperatures. In an effort to
understand these, we have since carried out additional fits to our data, which
we present here along with our original approach. In addition, an improved
\emph{Swift} calibration has given us more confidence in the modelling of the
UV-optical spectral energy distribution. Our new results agree more closely
with \cite{kulkarni+vankerkwijk10-1} than our original analysis, but
also reveal uncertainties in the system properties which make it
impossible to establish securely such fundamental properties as whether the
system is super-Chandrasekhar or not. We begin with a description of our
observational material.

\section{Observations and Reduction}

We obtained 140 spectra in the $B$ and $R$ bands covering the Balmer series on
the William Herschel Telescope in La Palma using the ISIS double-beam
spectrograph.  The first 22 spectra were taken in service mode on the night of
April 20, 2008, while the final 118 were acquired during a four night run
spanning April 29 to May 2, 2009. The data were debiassed, flat-fielded and
optimally extracted \citep{Horne:optimal,Marsh:optimal}.  The wavelength scale
for each spectrum was linearly interpolated in time from pairs of bracketing
arc calibration spectra. In 2008, the night was clear with seeing $0.8$ to
$1.6$ arcsec, however the Moon was full. In 2009, the Moon was only 40\%
illuminated, but conditions were poorer, and we lost a total of one night 
to clouds while the seeing varied from $1.0$ to $2.5$ arcsec. 

In addition to the optical spectroscopy, we obtained ultraviolet flux
measurements of \target\ with the \emph{Swift} satellite and accessed
additional archival data from 7 epochs of observations covering the interval
Aug 24 2009 -- Jan 13 2010. Our summed images had effective exposure times
ranging from 818 to 5608\,s. \target\ was well detected in all four UV filters
provided by \emph{Swift}/UVOT \citep{Swift:UVOT}. Aperture photometry with a
5'' extraction region and an annular sky region was performed using the SWIFT
data analysis package V3.7 within the HEASoft V6.10 release. Associated
zero-point calibration files and filter effective area curves (Jan 2011
release) were used to calculate the fluxes for each filter.

\section{Results}

In all that follows we will refer to the component most obvious in the spectra
as the primary star, and its companion as the secondary star. This means that 
the primary star is low mass and cool, while the secondary star is high mass and hot.
 
\subsection{The primary star}

Orbital motion of large amplitude was immediately apparent in the spectra. The
left panel of Fig.~\ref{fig:pbin_trail} shows phase-folded trailed spectra of
the first four Balmer lines displaying high-amplitude orbital motion and an
apparently single-lined binary star.

\begin{figure}
 \centering
\hspace*{\fill}
\includegraphics[width=0.65\columnwidth,angle=270]{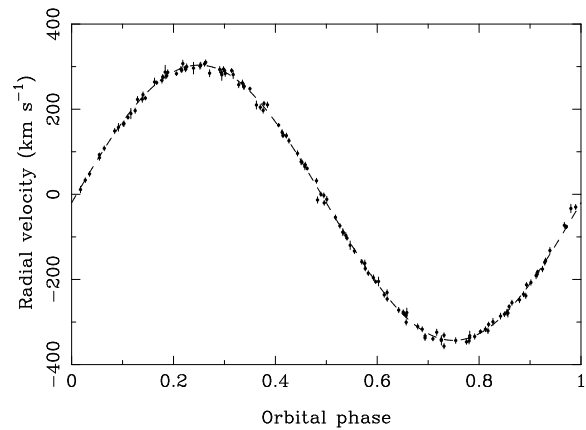}
\hspace*{\fill}
\caption{Radial velocities of H$\alpha$ from 2008 and 2009 folded on the ephemeris of
  this paper. \label{fig:halpha_rv}}
\end{figure}

We fitted the radial velocities (Fig.~\ref{fig:halpha_rv}) of the Balmer lines using combinations of 3
Gaussians for each line, with full width half maxima (FWHM) and depths
optimised to minimise $\chi^2$.
Given the large radial velocity amplitude of \target, the four night run in
2009 was sufficient to extrapolate back to the service night of a year earlier
to find a unique alias, leading to the following ephemeris (on a UTC timescale
corrected for light-travel time to the heliocentre):
\begin{equation}
 HJD = 2454846.17470 (8) + 0.18979154 (9) E,
\end{equation}
where the time of zero phase corresponds to the time when the primary star
is closest to Earth. While our period is consistent with that of
\cite{Badenes:SDSS1257} ($P = 0.18979(3)$ days), our zero phase is not, since 
on our ephemeris their zero phase occurs at  $E = 115.619 \pm 0.004$; this is 
the result of an error of $0.5$ days in the calculation of Julian Days
together with no allowance for light travel time in \cite{Badenes:SDSS1257} 
(C.Badenes, priv.\ comm.); we have in addition confirmed that our ephemeris 
correctly predicts the velocities of the SDSS spectra which were taken in 2003.

\begin{figure}
 \centering
\hspace*{\fill}
\includegraphics[width=0.65\columnwidth,angle=270]{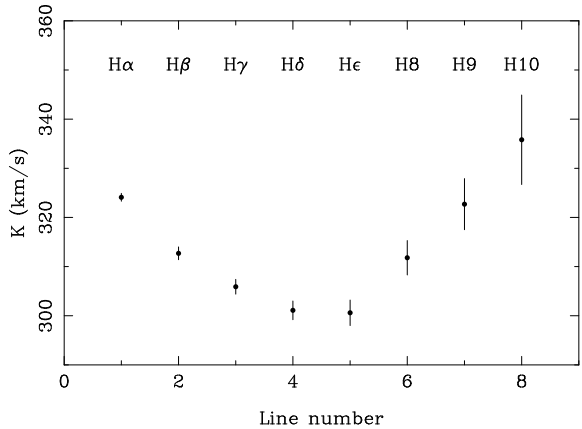}
\hspace*{\fill}
\caption{The radial velocity semi-amplitude of \protect\target\ from H$\alpha$
  to H10 showing the varying influence of the companion
  (which always reduces the amplitude) across the Balmer series.
\label{fig:k_vs_line}}
\end{figure}

Having established a unique period alias, from now on we use only the 2009
data because of differences in the instrumental setups between 2008 and 2009
which make consistent continuum fitting difficult. Sinusoidal fits were made
to the radial velocities measured for each Balmer line from H$\alpha$ to H10. The
semi-amplitudes were seen to vary from $324.1\pm 0.8\,\kms$ for H$\alpha$ to
$300.6\pm 2.6\,\kms$ for H$\epsilon$, rising again in the higher Balmer lines
(Fig.~\ref{fig:k_vs_line}).
The variation from line to line is a symptom of the presence of a second
broad-lined white
dwarf. As indicated earlier, and as will be seen later, the secondary white dwarf
is hotter and more massive than the primary. Therefore, owing to its temperature on the
one hand, and its high gravity on the other, the secondary is expected to contribute
relatively weakly at H$\alpha$ and in the higher Balmer series, but more
strongly in intermediate series lines. This is exactly what is seen in 
Fig.~\ref{fig:k_vs_line}, with the radial velocity semi-amplitude acting as a
proxy for the relative line strength from the two components. 

\subsection{The secondary star}

To search for more direct signs of a companion we first fitted and normalised
the continua of the spectra between the Balmer lines. Next we shifted out the
primary star's orbital motion and subtracted the average of the resulting
spectra. If only one component contributes significantly, this procedure
should leave only noise. Any contribution from a secondary star will cause
correlations amongst the residuals. The right-hand panel of
Fig.~\ref{fig:pbin_trail} shows such residuals which moreover are anti-phased
compared to the left-hand panel. This is the signature of a second
contributing white dwarf, albeit distorted by the subtraction process.

\begin{figure}
 \centering
\includegraphics[width=0.9\columnwidth]{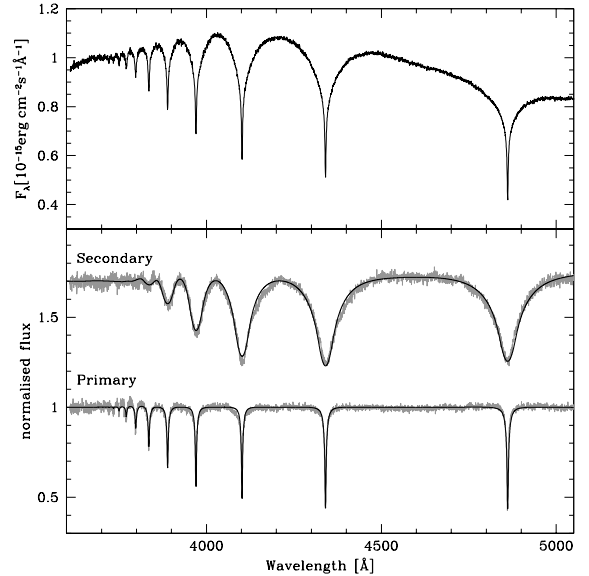}
\caption{The average $B$-band spectrum (top) and results of disentangling the
  spectra of \target\ (gray lines) along with model
  atmospheres (primary: $T_\mathrm{eff}=6900$\,K, $\log g=6.75$; secondary:
  $T_\mathrm{eff}=10300$\,K, $\log g=8.70$) for a light ratio of $0.7$
  (section~\protect\ref{sec:specfits}) and $v\sin i=1500\,\mathrm{km\,s^{-1}}$
  (black lines).  For disentangling, the radial-velocity semi-amplitudes
  were fixed at $K_1 = 330\,\kms$ and $K_2 = 100\,\kms$, but the results were
  insensitive to the precise value of $K_2$.
\label{fig:dtangle_b}}
\end{figure}
\begin{figure}
 \centering
\includegraphics[angle=270,width=0.85\columnwidth]{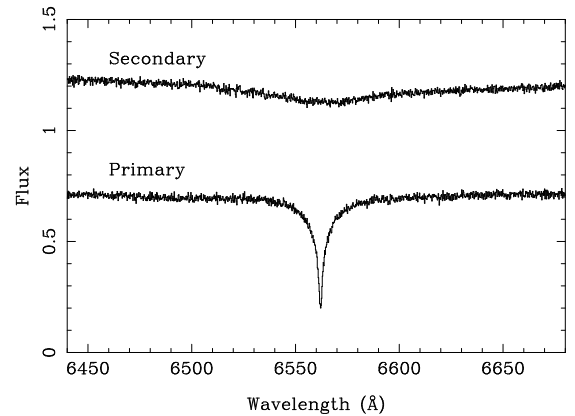}
\caption{The disentangled H$\alpha$ spectra with the secondary star
  component plotted at the top. As with Fig.~\protect\ref{fig:dtangle_b}, the
  continuum levels are indeterminate.
\label{fig:dtangle_r}}
\end{figure} 

The spectra of two stars within a binary can also be recovered using a
method applied successfully to main-sequence binaries known as 
``disentangling'' \citep{Sturm:disentangle}. Although the
broad absorption lines of white dwarfs make it harder to establish the reliable
normalisation required for disentangling, we tried it out to see if
any separation of the two spectra was possible. Its appeal is that it requires
no \emph{a priori} assumptions concerning the spectra of the two components,
although it cannot return the individual continua, and takes as input
continuum-normalised spectra.
The results for the blue spectra, plotted in Figs~\ref{fig:dtangle_b},
and over-plotted with model atmospheres, show a pair of low- and
high-gravity white dwarf spectra. The high-gravity secondary contributes
the broad line wings and it is thus clear why \cite{Badenes:SDSS1257} over-estimated 
the mass of the primary and had difficulties obtaining good spectral fits.
The process of disentangling leaves the continuum levels undetermined; we
discuss the model atmosphere fits further in section~\ref{sec:specfits}. The
disentangled H$\alpha$ profiles (Fig.~\ref{fig:dtangle_r})
are remarkable for an absence of any sharp line core in the high gravity
spectrum. H$\alpha$ lines in white dwarfs of this temperature usually display
a sharp dip at the line centre 2 or 3\AA\ in width as the result of non-LTE
effects \citep{Greenstein:1973,KoesterHerrero1988,Heber:1997}, but Fig.~\ref{fig:dtangle_r}
shows no sign at all of this. The figure also makes it clear 
why H$\alpha$ relatively faithfully traces the motion of the primary.

\subsection{Mass constraints}
We were unable to obtain a reliable direct measurement of the semi-amplitude
of the secondary star $K_2$ from our data, a consequence, we believe, of its
high gravity and the absence of any sharp line cores, combined with 
the difficulty of defining reliable continua.  However, when trying to 
obtain simultaneous fits to both stars, we found that the extra freedom 
of a secondary component always allowed
$K_1$ to increase above the directly-fitted values to a value consistently
close to $330 \pm 2 \,\kms$ which we take to be the true value of $K_1$.
The directly measured values are biassed towards lower values by the
contribution from the secondary, although as Fig.~\ref{fig:k_vs_line}
demonstrates, the effect is relatively slight in H$\alpha$ and in the higher 
Balmer series where the secondary star's high gravity ensures that it contributes little.
The value of $K_1$, along with the presence of a second
white dwarf leads to the mass constraints plotted in Fig.~\ref{fig:m1m2}.

\subsection{Spectroscopic fits}
\label{sec:specfits}

Our WHT data reveal that the narrow Balmer lines in \target\ extend up
to H12 (Fig.~\ref{fig:dtangle_b}), implying unambiguously that the
cooler white dwarf (primary) is of an unusually low surface gravity. In the
original submission of this paper, we modelled our WHT spectroscopy of
\target\ as we describe in Sect.\,~\ref{sec:specfits_dis}, finding that
\target\ is composed of a very low-mass, cool white dwarf plus a
hotter and rather massive white dwarf that is rapidly rotating. While
our paper was under review, \citet{kulkarni+vankerkwijk10-1} published an
independent study of \target, coming to broadly similar conclusions,
but with noticeable differences in the detailed white dwarf
parameters. In the light of this development, we re-visited the
modelling of the WHT spectroscopy and the ultraviolet (UV)/optical
spectral energy distribution (SED), adopting three different approaches which
we now detail.

\subsubsection{Disentangled spectra} 
\label{sec:specfits_dis}

We modelled the two individual spectral components resulting from the
disentangling of Fig.~\ref{fig:dtangle_b} using the fitting routine of
\citet{rebassa-mansergasetal07-1}. The grid of DA model spectra which was
calculated as in \citet{koesteretal05-1}, includes the updated Stark
broadening of the Balmer lines of \citet{tremblay+bergeron09-1}, and
spans $5< \log g<9.5$ in steps of 0.25 and 131 temperatures sampling
the range $6000 <T_\mathrm{eff}< 100000\,\mathrm{K}$ in an
optimised way.

\begin{figure}
 \centering
\includegraphics[width=0.7\columnwidth,angle=270]{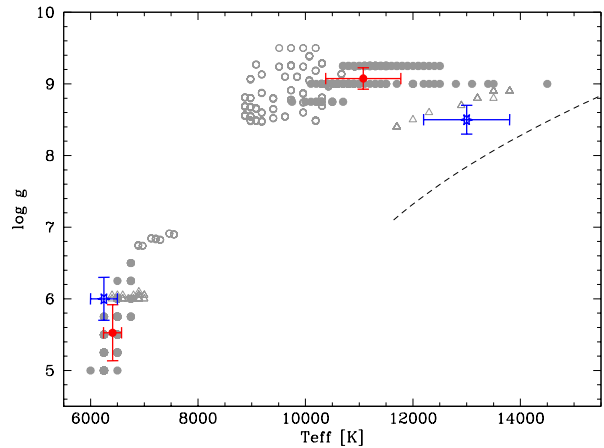}
\caption{\label{fig:wdfit_summary} Constraints on the effective
  temperatures and surface gravities of the two white dwarfs in
  \target. Fits to the disentangled spectra
  (Sect.\,\ref{sec:specfits_dis}) were carried out with light ratios
  0.7--1.3 in steps of 0.1, and secondary star rotation rates $v\sin
  i=500-1750\,\mathrm{km\,s^{-1}}$ in steps of
  $250\,\mathrm{km\,s^{-1}}$, and are shown as open circles. The
  best-fit parameters from fitting the phase-resolved 2009 WHT
  spectroscopy (Sect.\,\ref{sec:specfits_trail}) are shown as filled
  circles, their averages and standard deviations are indicated by the
  red error bars ($T_\mathrm{1,eff}=6341\pm139$\,K, $\log
  g_1=5.40\pm0.37$, $T_\mathrm{2,eff}=11111\pm704$\,K, and $\log
  g_2=9.13\pm0.14$). Fits to the SDSS/UVOT broad-band photometry are
  shown as triangles (Sect.~\ref{sec:specfits_sed}). The blue error
  bars show the best-fit parameters of \citet{kulkarni+vankerkwijk10-1}, and the dashed line
  indicates the occurrence of maximum Balmer-line strength. }
\end{figure}

\begin{table}
\caption{\label{tab:summary} Summary of effective temperature and gravity measurements for \protect\target.}
\begin{tabular}{lr@{$\pm$}lr@{$\pm$}lr@{$\pm$}lr@{$\pm$}l}
\hline\hline
Source & 
\multicolumn{2}{c}{$T_{\mathrm{eff},1}$} & 
\multicolumn{2}{c}{$\log g_1$} & 
\multicolumn{2}{c}{$T_{\mathrm{eff},2}$} & 
\multicolumn{2}{c}{$\log g_2$} \\ 
& 
\multicolumn{2}{c}{[K]} & 
\multicolumn{2}{c}{[$\mathrm{cm}\,\mathrm{s}^{-2}$]} & 
\multicolumn{2}{c}{[K]} & 
\multicolumn{2}{c}{[$\mathrm{cm}\,\mathrm{s}^{-2}$]} \\
\hline
Disentangling  & 
$7200$ & $350$ & $6.9$ & $0.1$ & $9800$  & $1000$ & $9.0$ & $0.4$ \\
Phase-resolved &
$6340$ & $140$ & $5.4$ & $0.4$ & $11100$ & $710$   & $9.1$ & $0.2$ \\
UV-optical SED &
$6620$ & $200$ & $6.1$ & $0.1^a$ &
$12400$ & $710$ & $8.6$ & $0.2$ \\
KvK2010  & 
$6250$ & $250$ & $6.0$  & $0.3^a$ 
& $13000$ & $800$ & $8.5$ & $0.2$ \\
\hline
\end{tabular}\\
$^a$Fits limited to $\log g > 6.0$.
\end{table}

\begin{figure*}
 \centering \includegraphics[width=0.6\textwidth,angle=270]{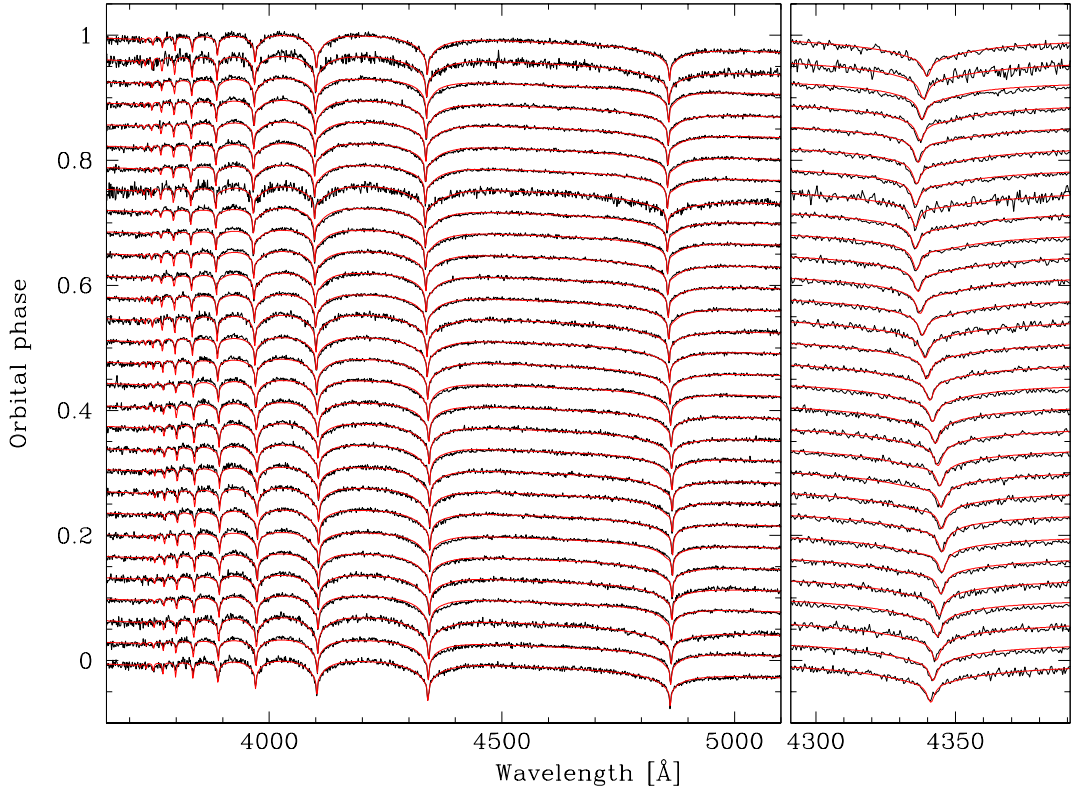}
\caption{\label{fig:wdfit_trail} Composite model fits (red) to the
  2009 WHT spectroscopy of \target, averaged in 30 orbital phase bins
  (black). Each individual spectrum was fitted independently,
  resulting in 30 solutions for $T_\mathrm{1,eff}$, $\log g_1$,
  $T_\mathrm{2,eff}$, $\log g_2$ (see Fig.\,\ref{fig:wdfit_summary}),
  and the relative flux contribution of both white dwarfs.}
\end{figure*}

Given that the relative flux ratio of the disentangled components is
indeterminate, we carried out fits adopting a range of ``light ratios'',
$l=0.7-1.3$ (in steps of 0.1), defined such that the secondary star
contributes $l$ while the primary contributes $2-l$ to the observed
light. Changing the light ratio modifies the ratio of the Balmer line
equivalent widths between the two components, and light ratios outside the
adopted range resulted in unphysical line strengths for one or both
components. In addition, while spectral disentangling adopts a single value
for the light ratio across the entire spectral range that is analysed, the
true light ratio of the two white dwarfs in \target\ is a function of the
wavelength. This may introduce systematic uncertainties (see
Sect.\,\ref{sec:specfits_trail}), although one would hope that the range of
light ratios adopted would reveal such systematics through variation in the
best fit parameters.  Conditions during our observations precluded any
reliable flux calibration of our spectra and we use no information from the
continuum.  For each assumed light ratio $l$, the model fit to the primary
results in a best-fit effective temperature and surface gravity, an example
for $l=0.7$ is shown in Fig.\,~\ref{fig:dtangle_b}.
Figure\,\ref{fig:wdfit_summary} 
illustrates that regardless of $l$, the
parameters of the primary are constrained within a narrow range,
$T_\mathrm{eff,1}=7200\pm350$\,K and $\log g_1=6.85\pm0.10$. This temperature
and gravity implies a very low mass for the primary, $M_1 = 0.20 \pm
0.05\,\msun$ \citep{paneietal07-1}. These fit parameters and others discussed
in this paper, including those of \cite{kulkarni+vankerkwijk10-1} listed as
KvK2010, are summarised in Table~\ref{tab:summary}.

\begin{figure}
 \centering
\includegraphics[width=0.6\columnwidth,angle=270]{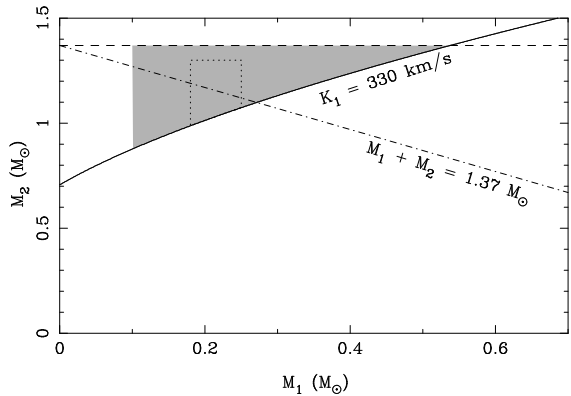}
\caption{Constraints upon the masses of the two stars in \target.  The
  solid curved line is the lower limit upon $M_2$ given $K_1 = 330
  \,\kms$.  The value of $M_2$ is bounded from above by the
  Chandrasekhar limit (horizontal dashed line), $M_\mathrm{Ch} = 1.37\,\msun$, 
  while a conservative minimum white dwarf mass of $0.1\,\msun$ places a lower limit upon
  $M_1$. These three constraints lead to the shaded triangular
  region. The dotted-line box outlines the more arguable region of parameter
space favored by the spectroscopic and evolutionary arguments outlined in 
section~\protect\ref{sec:present}.
\label{fig:m1m2}}
\end{figure}

The broad Balmer lines of the secondary (particularly H$\alpha$) suggest rapid
rotation. Assuming this, we found that acceptable fits were only obtained for
projected equatorial rotation speeds in the range $v\sin i = 500$ --
$1750\,\mathrm{km\,s^{-1}}$. The distributions of the best-fit parameters in
the $T_\mathrm{eff}-\log g$ plane are again illustrated in
Fig.\,\ref{fig:wdfit_summary}, with higher values of $v\sin i$ requiring lower
$\log g$ to reproduce the observed broad line profiles. These span a range
$T_{\mathrm{eff},2} = 9800\pm1000$\,K and $\log g_2=9.0\pm0.4$, corresponding
to $M_2 = 1.2 \pm0.2\,\msun$, consistent with the kinematic constraints of
Fig.~\ref{fig:m1m2}. These parameters broadly agree with the findings of
\citet{kulkarni+vankerkwijk10-1}, however, they found an even lower gravity
for the low-mass white dwarf primary ($\log g_1=6.0\pm0.3$), and a higher
temperature for the more massive secondary
($T_{\mathrm{eff},2}=13000\pm800$\,K). We will see below, as is also clear
from Table~\ref{tab:summary}, that the \textit{Swift} UVOT fluxes of
\target\ (section~\ref{sec:specfits_sed}) indeed suggest that
$T_{\mathrm{eff},2}$ from modelling the disentangled spectra is too low.

\subsubsection{Phase-binned flux spectra}
\label{sec:specfits_trail}

An assumption inherent to the spectral disentangling is a constant light
ratio, i.e. relative flux contribution of both components, across the
wavelength range under analysis. Our application of the method over the entire
wavelength range of the blue WHT spectra violates this assumption, which will
result in systematic errors in the relative strengths of the individual Balmer
lines, although as we remarked, our use of a range of light ratios might be
expected to transfer any resulting variation into the final uncertainties.
However, as an alternative we decided to carry out more direct fits to the WHT
spectra. We binned the 118 individual spectra into 30 phase bins, and fitted
each of them separately with the sum of two white dwarf model spectra drawn
from the same model grids as in Sect.\,\ref{sec:specfits_dis}. Observed and
model spectra were normalised prior to the fits. The fits were repeated for a
range of rotational velocities for the more massive white dwarf, $500 < v \sin
i < 1750\,\kms$, in steps of $250\,\kms$.

Figure~\ref{fig:wdfit_trail}
 shows the independent best-fits to the 30
phase-binned spectra for $v\sin i=1000$\,km/s.  The distributions of the 30
individual best-fit white dwarf parameters are shown in
Fig.~\ref{fig:wdfit_summary}. Each of the 30 independent fits 
leads to a pair of points in this figure, one in each of the two separate
groups of points. Averaging these groups leads to 
$T_\mathrm{eff,1}=6340\pm140$\,K, $\log g_1=5.40\pm0.37$,
$T_\mathrm{eff,2}=11100\pm710$\,K, and $\log
g_2=9.13\pm0.14$. Higher (lower) values for $v \sin i$ shifts both
white dwarfs to lower (higher) $T_\mathrm{eff}$ and $\log g$,
by a few hundred degrees and 0.1-0.2\,dex, with no significant difference
in the quality of the fits. 

Compared to the fit of the disentangled spectra
(Sect.\,\ref{sec:specfits_dis}), the two white dwarfs move further apart in
the ($T_\mathrm{eff}, \log g$) plane, somewhat in the sense of
\citet{kulkarni+vankerkwijk10-1}. Interestingly, $\log g_1=5.40\pm0.37$ from
fitting the phase-binned spectra is even lower than that reported by
\citet{kulkarni+vankerkwijk10-1}, $\log g_1=6.0\pm0.3$. However, their
best-fit value for $\log g_1$ is on the boundary of their model grid, and thus
may have been limited by the range of their models. This
exceedingly low gravity by white dwarf standards raises evolutionary problems
as we elucidate in section~\ref{sec:present}.

The best-fit parameters show considerable scatter in the gravity of the
primary, $\log g_1$ and the temperature of the secondary,
$T_{\mathrm{eff},2}$, which are poorly constrained by the standards of
single white dwarf spectra of comparable quality. Presumably the simultaneous fit
of two spectra introduces an element of degeneracy not present for single
white dwarfs.

\subsubsection{SDSS and UVOT spectral energy distribution}
\label{sec:specfits_sed}

\begin{figure}
 \centering \includegraphics[width=0.7\columnwidth,angle=270]{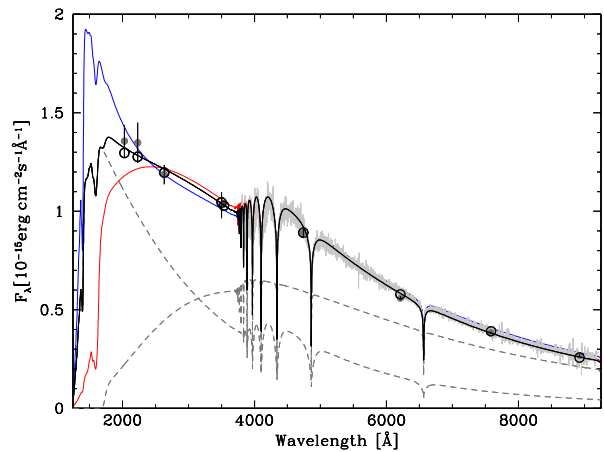}
\caption{\label{fig:wdfit_sed} The spectral energy distribution of
  \target. SDSS and UVOT broad-band fluxes are shown as filled circles, also
  plotted is the SDSS spectrum (gray). The best-fit to the broad-band fluxes
  is shown by the solid black line (middle in the FUV), with the individual
  contributions of the cool and hot components by dashed gray lines. 
  A composite model for the parameters of \citet{kulkarni+vankerkwijk10-1} is
  shown in blue (highest in the FUV); our best fit to the
  phase-dependent WHT spectra (Sect.~\ref{sec:specfits_trail},
  Fig.\,~\ref{fig:wdfit_trail}) in red (lowest in the FUV).}
\end{figure}

At optical wavelengths the cool, low-mass primary dominates the
observed flux of \target. However, given the substantial difference in
effective temperatures, the flux contribution of the two white dwarfs
is expected to reverse in the ultraviolet, which motivated our
\textit{Swift} TOO observations. These show an increasing flux towards
shorter wavelengths (Fig.\,\ref{fig:wdfit_sed}).
 We have fitted the
energy distribution spanned by the broad-band fluxes in the UVOT uvw2,
uvm2, uvw1, and b-bands and in the SDSS $ugriz$-bands with composite
white dwarf models, using the same model grid as in
Sect.\,\ref{sec:specfits_dis} and~\ref{sec:specfits_trail}. More
specifically, we fold all model spectra through the spectral response
curves of the \textit{Swift} and SDSS bandpasses, and calculate
absolute magnitudes using the cooling sequences of
\citet{bergeronetal95-2}\footnote{An updated 2006 version is available
  at http://www.astro.umontreal.ca/\~{}bergeron/CoolingModels}, which we
extrapolated down to $\log g=6.0$. In a final step, the co-added set
of absolute magnitudes are fitted to the \textit{Swift} and SDSS
data. The best fit formally gives 
$T_\mathrm{1,eff}=6620\pm200\,$K, $\log g_1=6.1\pm0.1$,
$T_\mathrm{2,eff}=12400\pm400\,$K, and $\log g_2=8.6\pm0.2$,
which are similar to those of \citet{kulkarni+vankerkwijk10-1}, but suffer
from the same limitation, namely being hard-bounded at $\log g=6.0$, in our
case by the cooling sequences we used. Nevertheless, the UVOT fluxes clearly
favor a temperature for the secondary at the high end of the values returned
by our spectroscopic fits.

Figure\,\ref{fig:wdfit_sed} illustrates our average best-fit to the
phase-binned spectra, the fit to the UV/optical broad-band fluxes, and
a model computed using the parameters of Table\,1 in
\citet{kulkarni+vankerkwijk10-1} along with the \textit{Swift} and
SDSS broad-band fluxes, and the SDSS spectrum. The parameters used in the
models are those summarised in Table~\ref{tab:summary}. The models diverge in the far
ultraviolet and will be easily distinguishable with \emph{HST} spectra. The
model based upon the mean of our phase-binned data is too cool, while
that based upon \citet{kulkarni+vankerkwijk10-1} is a little too hot in the
FUV and too cool at optical wavelengths. The temperature of the
secondary from the disentangled spectra ($9800\,$K) is obviously much
too cool and we plot no corresponding SED in Fig.~\ref{fig:wdfit_sed}.

In summary, the optical and ultraviolet characteristics of \target\ are
consistent with a double-degenerate consisting of a cool, extremely low-mass
white dwarf plus a rather massive, hotter white dwarf. We estimate the
distance to the binary to be $d\sim100\,$pc, larger than the $48\,$pc
estimated by \cite{Badenes:SDSS1257} because of the larger size of the
primary.  However, the data currently available are insufficient to fully
constrain the five free parameters ($T_\mathrm{1,eff}$, $\log g_1$,
$T_\mathrm{2,eff}$, $\log g_2$ and $v \sin i$).

\section{Discussion}

\subsection{Present state of \target}
\label{sec:present}

\begin{figure}
 \centering \includegraphics[width=0.7\columnwidth,angle=270]{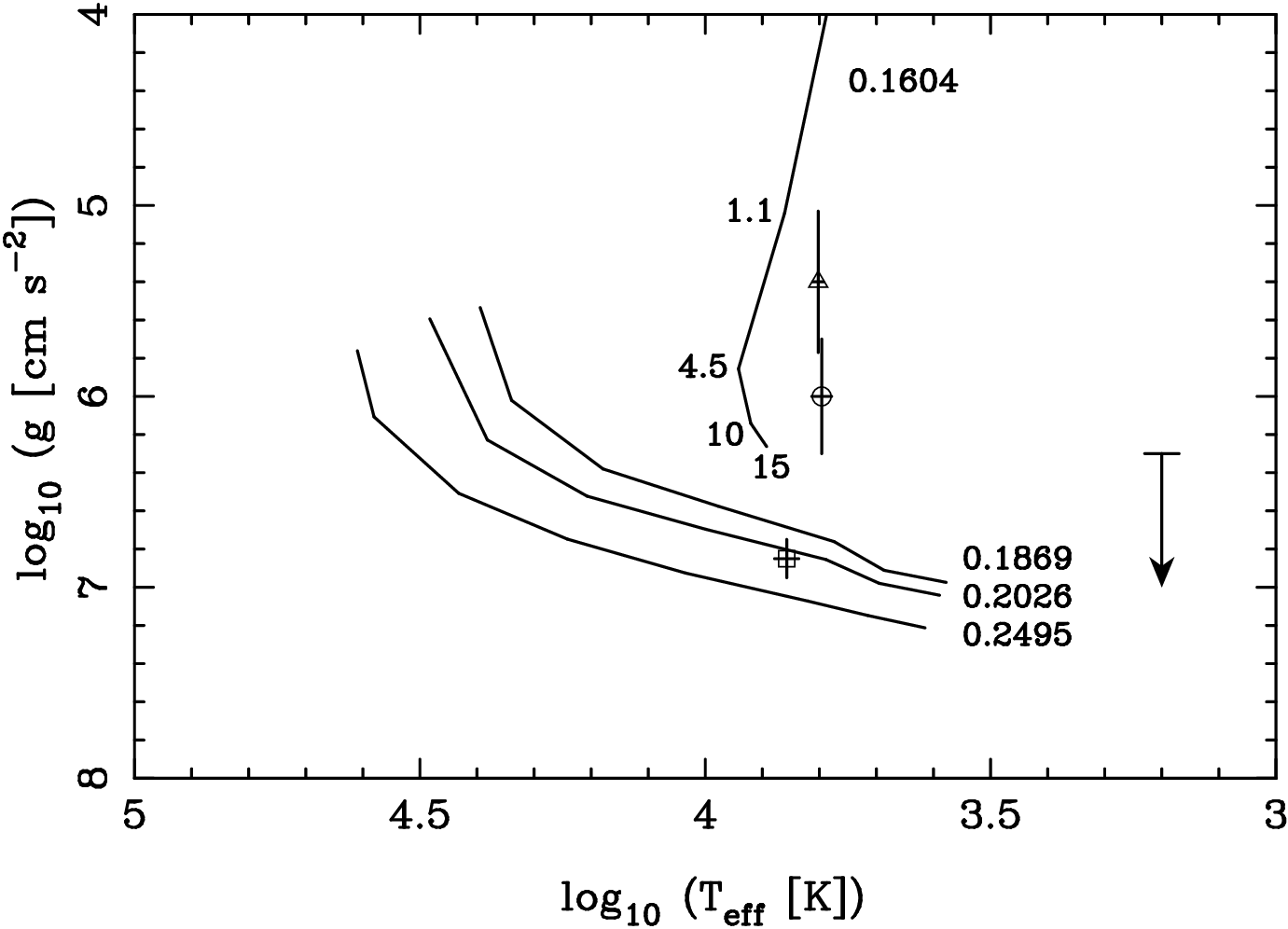}
\caption{\label{fig:logg_teff} From top to bottom the symbols with error bars
show the temperature and gravity of the low-mass primary based upon 
(i) our phase-resolved spectra (triangle), (ii)
\protect\cite{kulkarni+vankerkwijk10-1}'s equivalent result
(circle), and (iii) our fit to the disentangled spectra (square).
The lines show evolutionary models for different mass helium 
white dwarfs from \protect\cite{paneietal07-1}. The cooling ages in Gyr are
labelled on the $0.1604\,\msun$ model which is slowed by hydrogen fusion. 
The arrow in the lower right indicates the lower limit on gravity based upon
assuming that the secondary contributes at least 25\% of the flux.}
\end{figure}

The precise nature of the low mass primary star in \target\ presents us with a
puzzle.
As Fig.~\ref{fig:logg_teff}
shows, the fits to the disentangled spectra are consistent with a mass of 
around $0.2\,\msun$ \citep{paneietal07-1}. The cooling age in this case
is $\sim 1\,$Gyr which just about allows the secondary star (cooling age 1 to $2\,$Gyr, for CO
and ONe models \citep{bergeronetal95-2, althaus:2007}) to have a longer
cooling age than the primary, as we expect, since the massive white dwarf
presumably formed before the helium white dwarf.  In contrast, our phase-resolved
fits, which fit the data directly, as well as the similar fits of
\cite{kulkarni+vankerkwijk10-1}, suggest a much lower gravity and a mass $<
0.16 \, \msun$. In fact, as both Figs~\ref{fig:wdfit_summary} and
\ref{fig:logg_teff} show, the gravity of the primary is very poorly
constrained by the spectroscopic fits, ranging from $\log g = 5$ to $6.6$ (and
even then the lower bound is set by the model grid). Crucially, this range
allows the primary star to sit on either side of the division at $\sim
0.18\,\msun$ below which residual hydrogen on the surface of helium white
dwarfs can maintain fusion and greatly slow their cooling whereas above,
hydrogen shell flashes can remove the hydrogen and allow normal cooling
\citep{Kippenhahnetal1968, Webbink1975, Driebe:hewds, Sarnaetal2000,
  Althausetal2001, Nelsonetal2004, paneietal07-1}. As Fig.~\ref{fig:logg_teff}
shows, if the mass of the primary is below this limit, then in order to keep
the cooling age of the primary short enough to allow the secondary to be as
hot as it is, the gravity must lie at the lower end of the range with $\log g
\approx 5$.

Such a low gravity is hard to believe because, for reasonable
assumptions about the masses of the two stars, the primary becomes so large
that its drowns out the light from the secondary. Quantitatively, the ratio 
$r$ of the flux from the secondary divided by 
that from  the primary, at wavelength $\lambda$ is given by
\begin{equation}
r = \frac{R_2^2 F(\lambda, T_2)}{R_1^2 F(\lambda, T_1)},
\end{equation}
where $F$ is the astrophysical flux density (flux per unit wavelength per unit
area of the star). Using $R^2 \propto M/g$ and applying the observed 3:1
maximum ratio (from the disentangling light ratio limits) gives therefore
\begin{equation}
r = \frac{M_2 g_1 F(\lambda, T_2)}{M_1 g_2 F(\lambda, T_1)} > \frac{1}{3} .
\end{equation}
This then leads to a lower limit upon the gravity of the primary:
\begin{equation}
g_1 > \frac{1}{3} \frac{M_1 g_2 F(\lambda, T_1)}{M_2 F(\lambda, T_2)}.
\end{equation}
Assuming $M_2 = 1.0\,\msun$, $\log g_2 = 8.64$, $M_1 = 0.15$ (a lower limit to
give the least stringent constraint on $g_1$), $T_1 = 6300$, $T_2 = 14000\,$K
(an upper limit, again to give the weakest constraint on $g_1$), and taking
the ratio of model atmospheres at $\lambda = 483\,$nm ($g$-band central wavelength), we find
$\log g_1 > 6.3$. We plot this lower limit on the right-hand 
side of Fig.~\ref{fig:logg_teff}. This limit suggests that the surface gravity
of the primary star is at the upper end of the range returned by the
spectroscopic fits. If the primary has a mass below $0.18\,\msun$, as the
spectroscopic fits seem to favour, then the evolutionary models suggest that
its cooling age is of order $10\,$Gyr.

We are left with a choice between two unpalatable alternatives: either the
primary has a mass above $> 0.18\,\msun$, as suggested by the fits to the
disentangled spectra but in contradiction with both our direct fits and those
of \cite{kulkarni+vankerkwijk10-1}, or it has a mass below this limit and is
therefore extremely old, making the youth of the more massive secondary
impossible to understand, since it should have formed first.

We do not have a satisfactory resolution of this problem, but marginally
prefer the higher mass solution ($M_1 \sim 0.2\,\msun$) for the primary to
avoid the relative age problem. This supposes that the evolutionary models are
correct; further work to fix $M_1$ is encouraged.  If we are right that the
primary mass in \target\ is $0.2\,\msun$, then the mass of the secondary is $>
1\,\msun$ (Fig.~\ref{fig:m1m2}). This puts the secondary close to the dividing
line between CO and ONe white dwarfs which is thought to lie at around
$1.05\,\msun$ for solar metallicity progenitors, corresponding to an initial
mass of about $5.5\,\msun$ \citep{Meng:IFMR}.

We suspect that the discrepancies in the best-fit parameters between the
various fits result from difficulties in establishing reliable continua given
the extremely broad wings from the high mass white dwarf and, in our case at
least, wavelength dependent slit losses during the rather poor conditions of
our observing run.  Differences between the model atmospheres and data are
visible in both Figs~\ref{fig:dtangle_b} and \ref{fig:wdfit_trail}. As a
result, our best-fit parameters suffer from some systematic errors, which are
only compounded by the degeneracy from fitting two spectra. The independent
evidence provided by the UV-optical SED favors a high temperature for the
secondary as found by \cite{kulkarni+vankerkwijk10-1}, suggesting that their
data are less affected by continuum problems than ours, although their low
gravity for the primary possibly indicates that they are not entirely immune
to the same problems. Further observations are needed to clarify this issue.

\subsection{Anomalous broadening}
Although there has been debate over the first phase of mass transfer in the
formation of double white dwarfs, the second phase, involving mass transfer
onto the white dwarf that forms first, is almost universally assumed to lead to
the formation of a common envelope \citep{Han:DDs,Nelemans:DWDs}. However, the
absence of a narrow core in H$\alpha$ of the secondary white dwarf in
\target\ is suggestive of rapid rotation with $v \sin i > 500\,\kms$, which
throws doubt upon this assumption since we estimate from consideration of the
moment of inertia of the white dwarf and the angular momentum of the accreted
material that it should have accreted $> 0.02\,\msun$ to attain this rotation
rate. This would require a much more prolonged period of accretion than allowed by
a common envelope, given a maximum accretion rate onto a white dwarf of
a few $\times 10^{-7}\,\msun\,\mbox{yr}^{-1}$ \citep{Nomoto:wdaccretion2}. This is
reminiscent of evolutionary scenarios involving stable second phases of mass
transfer discussed by \cite{Sarna:DDs}. A possible near-relative of
\target\ is the star HD~49798 in which a massive, rapidly spinning white dwarf
orbits an sdO star \citep{Mereghetti:fastWD}. Uncertainty remains however
over the cause of the broad profiles. There is no sign of Zeeman splitting in
H$\alpha$ (Fig.~\ref{fig:dtangle_r}), but as \cite{kulkarni+vankerkwijk10-1}
suggest, a magnetic field could make the core more difficult to see, and, if
suitably distributed in strength across the surface, would not necessarily
lead to obvious splitting. In a dual rotation-plus-magnetic-field model, the
field could also act to increase the lever arm for accretion and hence
reduce the amount of material needed to spin the white dwarf up. Alternatively,
if the atmosphere is helium-dominated, we will see deeper into the star which
may allow Stark broadening to wipe out the H$\alpha$ core. Finally, the
temperature of the secondary star is such that it could lie within the ZZ~Ceti
pulsational instability strip. It is observed \citep{Koesteretal1998} that
ZZ~Ceti stars show weak cores in H$\alpha$, probably the result of motion in
their photospheres \citep{KoesterKompa2007}. It would be worth observing
\target\ photometrically for pulsations.

\subsection{Future evolution}
Close pairs of white dwarfs are common within our Galaxy, and their future
evolution has been the subject of much discussion, not least as potential
progenitors of Type~Ia supernovae. For \target, the first step is clear:
gravitational radiation will reduce the orbital period to the point at which
mass transfer starts. The orbital period at this point will be about
$1.5\,$minutes which will be reached in $\sim 3 \times 10^9\,$years. The
outcome of mass transfer is uncertain and our constraints allow any of three
possibilities which are (i) explosion as a Type~Ia supernova, (ii)
accretion-induced collapse to a millisecond pulsar and (iii) survival of the
onset of mass transfer as a hydrogen-deficient ultra-compact binary. We
slightly favor the third possibility on the grounds that the mass ratio $q =
M_1/M_2 \approx 0.2$ is extreme enough to allow stable mass transfer
\citep{Nelemans:AMCVn,Marsh:mdot}, while we suspect the total mass to be less
than the Chandrasekhar mass. \target\ is therefore likely to become a
semi-detached, accreting double white dwarf (AM~CVn star). The secondary star
will become the accretor. Since it is massive for a white dwarf, and as long
as it is not an ONe white dwarf, the system is a good candidate progenitor of
the sub-luminous ``.Ia'' supernovae discussed by \cite{Bildstenetal2007} and
possibly observed by \cite{Kasliwaletal2010}.

\section{Conclusion}
We find that the putative white dwarf--black-hole/neutron star binary, \target, is a
double white dwarf. \target\ is composed of a very low mass $\sim 0.2\,\msun$
white dwarf together with an extremely massive ($> 1\,\msun$) white
dwarf. As long as the massive white dwarf avoids accretion-induced collapse or
explosion, \target\ will evolve into a hydrogen-deficient accreting binary
star, but may later explode as a sub-luminous Type~Ia. The massive white dwarf
shows signs of rapid rotation ($v \sin i > 500 \,\kms$) which suggests
that the most recent phase of mass transfer might not have involved a common
envelope, contrary to current models of double white dwarf populations.
Some inconsistencies in the parameters of the two white dwarfs remain that
are probably caused by difficulties in fitting their blended spectra; further
observations are required to clarify these.

\acknowledgements{ We thank Sergio Campana for the use of his additional
  \emph{Swift} data. We made use of SIMBAD, maintained by the Centre
  Donn\'{e}es astronomiques de Strasbourg and the National Aeronautics and
  Space Administration's Astrophysics Data System.  TRM, BTG, DS and JS
  acknowledge the financial support of the UK's STFC.  Balmer/Lyman lines in
  the models were calculated with the modified Stark broadening profiles of
  \cite{tremblay+bergeron09-1} kindly made available by the authors.
}

\bibliographystyle{apj}

\begin{thebibliography}{40}
\expandafter\ifx\csname natexlab\endcsname\relax\def\natexlab#1{#1}\fi

\bibitem[{{Althaus} {et~al.}(2007){Althaus}, {Garc{\'{\i}}a-Berro}, {Isern},
  {C{\'o}rsico}, \& {Rohrmann}}]{althaus:2007}
{Althaus}, L.~G., {Garc{\'{\i}}a-Berro}, E., {Isern}, J., {C{\'o}rsico}, A.~H.,
  \& {Rohrmann}, R.~D. 2007, \aap, 465, 249

\bibitem[{{Althaus} {et~al.}(2001){Althaus}, {Serenelli}, \&
  {Benvenuto}}]{Althausetal2001}
{Althaus}, L.~G., {Serenelli}, A.~M., \& {Benvenuto}, O.~G. 2001, \mnras, 324,
  617

\bibitem[{{Badenes} {et~al.}(2009){Badenes}, {Mullally}, {Thompson}, \&
  {Lupton}}]{Badenes:SDSS1257}
{Badenes}, C., {Mullally}, F., {Thompson}, S.~E., \& {Lupton}, R.~H. 2009,
  \apj, 707, 971

\bibitem[{{Bergeron} {et~al.}(1995){Bergeron}, {Wesemael}, \&
  {Beauchamp}}]{bergeronetal95-2}
{Bergeron}, P., {Wesemael}, F., \& {Beauchamp}, A. 1995, \pasp, 107, 1047

\bibitem[{{Bildsten} {et~al.}(2007){Bildsten}, {Shen}, {Weinberg}, \&
  {Nelemans}}]{Bildstenetal2007}
{Bildsten}, L., {Shen}, K.~J., {Weinberg}, N.~N., \& {Nelemans}, G. 2007,
  \apjl, 662, L95

\bibitem[{{Bragaglia} {et~al.}(1990){Bragaglia}, {Greggio}, {Renzini}, \&
  {D'Odorico}}]{Bragaglia:DDs}
{Bragaglia}, A., {Greggio}, L., {Renzini}, A., \& {D'Odorico}, S. 1990, \apjl,
  365, L13

\bibitem[{{Driebe} {et~al.}(1998){Driebe}, {Schoenberner}, {Bloecker}, \&
  {Herwig}}]{Driebe:hewds}
{Driebe}, T., {Schoenberner}, D., {Bloecker}, T., \& {Herwig}, F. 1998, \aap,
  339, 123

\bibitem[{{Eisenstein} {et~al.}(2006){Eisenstein}, {Liebert}, {Harris},
  {Kleinman}, {Nitta}, {Silvestri}, {Anderson}, {Barentine}, {Brewington},
  {Brinkmann}, {Harvanek}, {Krzesi{\'n}ski}, {Neilsen}, {Long}, {Schneider}, \&
  {Snedden}}]{Eisenstein:wds}
{Eisenstein}, D.~J., {Liebert}, J., {Harris}, H.~C., {Kleinman}, S.~J.,
  {Nitta}, A., {Silvestri}, N., {Anderson}, S.~A., {Barentine}, J.~C.,
  {Brewington}, H.~J., {Brinkmann}, J., {Harvanek}, M., {Krzesi{\'n}ski}, J.,
  {Neilsen}, Jr., E.~H., {Long}, D., {Schneider}, D.~P., \& {Snedden}, S.~A.
  2006, \apjs, 167, 40

\bibitem[{{Greenstein} \& {Peterson}(1973)}]{Greenstein:1973}
{Greenstein}, J.~L. \& {Peterson}, D.~M. 1973, \aap, 25, 29

\bibitem[{{Han}(1998)}]{Han:DDs}
{Han}, Z. 1998, \mnras, 296, 1019

\bibitem[{{Heber} {et~al.}(1997){Heber}, {Napiwotzki}, \& {Reid}}]{Heber:1997}
{Heber}, U., {Napiwotzki}, R., \& {Reid}, I.~N. 1997, \aap, 323, 819

\bibitem[{{Horne}(1986)}]{Horne:optimal}
{Horne}, K. 1986, \pasp, 98, 609

\bibitem[{{Iben} \& {Tutukov}(1984)}]{IbenTutukov1984}
{Iben}, Jr., I. \& {Tutukov}, A.~V. 1984, \apjs, 54, 335

\bibitem[{{Kasliwal} {et~al.}(2010){Kasliwal}, {Kulkarni}, {Gal-Yam}, {Yaron},
  {Quimby}, {Ofek}, {Nugent}, {Poznanski}, {Jacobsen}, {Sternberg}, {Arcavi},
  {Howell}, {Sullivan}, {Rich}, {Burke}, {Brimacombe}, {Milisavljevic},
  {Fesen}, {Bildsten}, {Shen}, {Cenko}, {Bloom}, {Hsiao}, {Law}, {Gehrels},
  {Immler}, {Dekany}, {Rahmer}, {Hale}, {Smith}, {Zolkower}, {Velur},
  {Walters}, {Henning}, {Bui}, \& {McKenna}}]{Kasliwaletal2010}
{Kasliwal}, M.~M., {Kulkarni}, S.~R., {Gal-Yam}, A., {Yaron}, O., {Quimby},
  R.~M., {Ofek}, E.~O., {Nugent}, P., {Poznanski}, D., {Jacobsen}, J.,
  {Sternberg}, A., {Arcavi}, I., {Howell}, D.~A., {Sullivan}, M., {Rich},
  D.~J., {Burke}, P.~F., {Brimacombe}, J., {Milisavljevic}, D., {Fesen}, R.,
  {Bildsten}, L., {Shen}, K., {Cenko}, S.~B., {Bloom}, J.~S., {Hsiao}, E.,
  {Law}, N.~M., {Gehrels}, N., {Immler}, S., {Dekany}, R., {Rahmer}, G.,
  {Hale}, D., {Smith}, R., {Zolkower}, J., {Velur}, V., {Walters}, R.,
  {Henning}, J., {Bui}, K., \& {McKenna}, D. 2010, \apjl, 723, L98

\bibitem[{{Kippenhahn} {et~al.}(1968){Kippenhahn}, {Thomas}, \&
  {Weigert}}]{Kippenhahnetal1968}
{Kippenhahn}, R., {Thomas}, H., \& {Weigert}, A. 1968, \zap, 69, 265

\bibitem[{{Koester} {et~al.}(1998){Koester}, {Dreizler}, {Weidemann}, \&
  {Allard}}]{Koesteretal1998}
{Koester}, D., {Dreizler}, S., {Weidemann}, V., \& {Allard}, N.~F. 1998, \aap,
  338, 612

\bibitem[{{Koester} \& {Herrero}(1988)}]{KoesterHerrero1988}
{Koester}, D. \& {Herrero}, A. 1988, \apj, 332, 910

\bibitem[{{Koester} \& {Kompa}(2007)}]{KoesterKompa2007}
{Koester}, D. \& {Kompa}, E. 2007, \aap, 473, 239

\bibitem[{{Koester} {et~al.}(2005){Koester}, {Napiwotzki}, {Voss}, {Homeier},
  \& {Reimers}}]{koesteretal05-1}
{Koester}, D., {Napiwotzki}, R., {Voss}, B., {Homeier}, D., \& {Reimers}, D.
  2005, \aap, 439, 317

\bibitem[{{Kulkarni} \& {van Kerkwijk}(2010)}]{kulkarni+vankerkwijk10-1}
{Kulkarni}, S.~R. \& {van Kerkwijk}, M.~H. 2010, \apj, 719, 1123

\bibitem[{{Marsh}(1989)}]{Marsh:optimal}
{Marsh}, T.~R. 1989, \pasp, 101, 1032

\bibitem[{{Marsh} {et~al.}(1995){Marsh}, {Dhillon}, \& {Duck}}]{Marsh:friends}
{Marsh}, T.~R., {Dhillon}, V.~S., \& {Duck}, S.~R. 1995, \mnras, 275, 828

\bibitem[{{Marsh} {et~al.}(2004){Marsh}, {Nelemans}, \& {Steeghs}}]{Marsh:mdot}
{Marsh}, T.~R., {Nelemans}, G., \& {Steeghs}, D. 2004, \mnras, 350, 113

\bibitem[{{Meng} {et~al.}(2008){Meng}, {Chen}, \& {Han}}]{Meng:IFMR}
{Meng}, X., {Chen}, X., \& {Han}, Z. 2008, \aap, 487, 625

\bibitem[{{Mereghetti} {et~al.}(2009){Mereghetti}, {Tiengo}, {Esposito}, {La
  Palombara}, {Israel}, \& {Stella}}]{Mereghetti:fastWD}
{Mereghetti}, S., {Tiengo}, A., {Esposito}, P., {La Palombara}, N., {Israel},
  G.~L., \& {Stella}, L. 2009, Science, 325, 1222

\bibitem[{{Napiwotzki} {et~al.}(2003){Napiwotzki}, {Drechsel}, {Heber}, {Karl},
  {Pauli}, {Christlieb}, {Hagen}, {Reimers}, {Koester}, {Moehler}, {Homeier},
  {Leibundgut}, {Renzini}, {Marsh}, {Nelemans}, \&
  {Yungelson}}]{Napiwotzki:SPY}
{Napiwotzki}, R., {Drechsel}, H., {Heber}, U., {Karl}, C., {Pauli}, E.-M.,
  {Christlieb}, N., {Hagen}, H.-J., {Reimers}, D., {Koester}, D., {Moehler},
  S., {Homeier}, D., {Leibundgut}, B., {Renzini}, A., {Marsh}, T.~R.,
  {Nelemans}, G., \& {Yungelson}, L. 2003, in NATO ASIB Proc. 105: White
  Dwarfs, 39

\bibitem[{{Nelemans} {et~al.}(2001{\natexlab{a}}){Nelemans}, {Portegies Zwart},
  {Verbunt}, \& {Yungelson}}]{Nelemans:AMCVn}
{Nelemans}, G., {Portegies Zwart}, S.~F., {Verbunt}, F., \& {Yungelson}, L.~R.
  2001{\natexlab{a}}, \aap, 368, 939

\bibitem[{{Nelemans} {et~al.}(2001{\natexlab{b}}){Nelemans}, {Yungelson},
  {Portegies Zwart}, \& {Verbunt}}]{Nelemans:DWDs}
{Nelemans}, G., {Yungelson}, L.~R., {Portegies Zwart}, S.~F., \& {Verbunt}, F.
  2001{\natexlab{b}}, \aap, 365, 491

\bibitem[{{Nelson} {et~al.}(2004){Nelson}, {Dubeau}, \&
  {MacCannell}}]{Nelsonetal2004}
{Nelson}, L.~A., {Dubeau}, E., \& {MacCannell}, K.~A. 2004, \apj, 616, 1124

\bibitem[{{Nomoto} {et~al.}(2007){Nomoto}, {Saio}, {Kato}, \&
  {Hachisu}}]{Nomoto:wdaccretion2}
{Nomoto}, K., {Saio}, H., {Kato}, M., \& {Hachisu}, I. 2007, \apj, 663, 1269

\bibitem[{{Panei} {et~al.}(2007){Panei}, {Althaus}, {Chen}, \&
  {Han}}]{paneietal07-1}
{Panei}, J.~A., {Althaus}, L.~G., {Chen}, X., \& {Han}, Z. 2007, \mnras, 382,
  779

\bibitem[{{Rebassa-Mansergas} {et~al.}(2007){Rebassa-Mansergas},
  {G{\"a}nsicke}, {Rodr{\'{\i}}guez-Gil}, {Schreiber}, \&
  {Koester}}]{rebassa-mansergasetal07-1}
{Rebassa-Mansergas}, A., {G{\"a}nsicke}, B.~T., {Rodr{\'{\i}}guez-Gil}, P.,
  {Schreiber}, M.~R., \& {Koester}, D. 2007, \mnras, 382, 1377

\bibitem[{{Robinson} \& {Shafter}(1987)}]{Robinson:DDs}
{Robinson}, E.~L. \& {Shafter}, A.~W. 1987, \apj, 322, 296

\bibitem[{{Roming} {et~al.}(2005){Roming}, {Kennedy}, {Mason}, {Nousek}, {Ahr},
  {Bingham}, {Broos}, {Carter}, {Hancock}, {Huckle}, {Hunsberger}, {Kawakami},
  {Killough}, {Koch}, {McLelland}, {Smith}, {Smith}, {Soto}, {Boyd},
  {Breeveld}, {Holland}, {Ivanushkina}, {Pryzby}, {Still}, \&
  {Stock}}]{Swift:UVOT}
{Roming}, P.~W.~A., {Kennedy}, T.~E., {Mason}, K.~O., {Nousek}, J.~A., {Ahr},
  L., {Bingham}, R.~E., {Broos}, P.~S., {Carter}, M.~J., {Hancock}, B.~K.,
  {Huckle}, H.~E., {Hunsberger}, S.~D., {Kawakami}, H., {Killough}, R., {Koch},
  T.~S., {McLelland}, M.~K., {Smith}, K., {Smith}, P.~J., {Soto}, J.~C.,
  {Boyd}, P.~T., {Breeveld}, A.~A., {Holland}, S.~T., {Ivanushkina}, M.,
  {Pryzby}, M.~S., {Still}, M.~D., \& {Stock}, J. 2005, Space Science Reviews,
  120, 95

\bibitem[{{Sarna} {et~al.}(2000){Sarna}, {Ergma}, \& {Ger{\v s}kevit{\v
  s}-Antipova}}]{Sarnaetal2000}
{Sarna}, M.~J., {Ergma}, E., \& {Ger{\v s}kevit{\v s}-Antipova}, J. 2000,
  \mnras, 316, 84

\bibitem[{{Sarna} {et~al.}(1996){Sarna}, {Marks}, \& {Connon
  Smith}}]{Sarna:DDs}
{Sarna}, M.~J., {Marks}, P.~B., \& {Connon Smith}, R. 1996, \mnras, 279, 88

\bibitem[{{Simon} \& {Sturm}(1994)}]{Sturm:disentangle}
{Simon}, K.~P. \& {Sturm}, E. 1994, \aap, 281, 286

\bibitem[{{Tremblay} \& {Bergeron}(2009)}]{tremblay+bergeron09-1}
{Tremblay}, P.-E. \& {Bergeron}, P. 2009, \apj, 696, 1755

\bibitem[{{Webbink}(1975)}]{Webbink1975}
{Webbink}, R.~F. 1975, \mnras, 171, 555

\bibitem[{{Webbink}(1984)}]{Webbink:DDs}
---. 1984, \apj, 277, 355

\end{thebibliography}

\end{document}